\pdfoutput=1

\documentclass{ws-ijmpb}
\usepackage{graphicx}
\usepackage{epsfig}
\usepackage{hyperref}

\begin{document}

\markboth{Freericks, Nikoli\'c and Frieder}
{The nonequilibrium quantum many-body problem as a \ldots}

%
\catchline{}{}{}{}{}
%

\title{THE NONEQUILIBRIUM QUANTUM MANY-BODY PROBLEM AS A PARADIGM FOR EXTREME DATA SCIENCE }

\author{J. K. Freericks}

\address{Department of Physiocs, Georgetown University, 37$^{\rm th}$ and O Sts. NW\\
Washington, District of Columbia 20057, U. S. A.\\
freericks@physics.georgetown.edu }

\author{B. K. Nikoli\'c}

\address{Department of Physics and Astronomy, University of Delaware, Newark, DE 19716, U. S. A.\\
bnikolic@udel.edu}

\author{O. Frieder}

\address{Department of Computer Science, Georgetown University, 37$^{\rm th}$ and O Sts. NW\\
Washington, District of Columbia 20057, U. S. A.\\
ophir@ir.cs.georgetown.edu }

\maketitle

\begin{history}
\received{Day Month Year}
\revised{Day Month Year}
\end{history}

\begin{abstract}
Generating big data pervades much of physics. But some problems, which we call extreme data problems, are
too large to be treated within big data science. The nonequilibrium quantum many-body problem on a lattice  is just such a problem,
where the Hilbert space grows exponentially with system size and rapidly becomes too large to fit on any computer (and can
be effectively thought of as an infinite-sized data set). Nevertheless, much progress has been made with computational methods on
this problem, which serve as a paradigm for how one can approach and attack extreme data problems. In addition, 
viewing these physics problems from a computer-science perspective leads to new approaches that can be tried
to solve them more accurately and for longer times. We review a number of these different ideas here.
\end{abstract}


\section{Extreme data science}

Conventional forms of big data science and data mining involve the creation of large finite data sets and the 
analysis of that data to produce insight and to predict future behavior of the system. Many physics problems
are being analyzed in this big data paradigm, such as high-energy-physics scattering experiments, which routinely
create petabytes of data. Other fields of science also involve big data science, with large datasets being analyzed in fields as diverse as weather forecasting, genomics, and neuroscience. Many algorithms and data analysis tools are used to interact with such datasets; often they are created specifically for a given data set, even if they employ more
general purpose algorithms and strategies for the data analysis.

We discuss physics problems for which the datasets are so large they can effectively be treated as
infinite in size, which we call {\it extreme data science}. In general, such problems are impossible to solve unless at any
given time, only a small subset of the large dataspace is physically active, and that the system maintains a reasonably small footprint within the data space for future times as well. 

One technique which has been developed in the previous decade for dealing with big data in computer science is the technique of compressive sensing. This method represents a multidimensional signal with a much lower dimensional model. Instead of employing conventional signal compression techniques, which aim to capture the whole signal, compressive sensing attempts to directly capture the sparse signal\cite{candes_romberg_tao,donobo}. One can understand how this works in two ways: (1) most data in real world problems are overdetermined  and hence can be represented in a  more sparse fashion and (2) from a mathematical standpoint, a time series originating from a Fourier series or from a Fourier transform look similar on a finite time interval, but the series can be represented sparsely, while the transform cannot. This latter feature is now being exploited in a wide range of physics problems.

The unprecedented growth of data has also challenged the most affordable computational procedures within big data science---namely numerical linear algebra. Even partial eigenvalue or singular value decompositions become too expensive for matrices employed in graph mining  of large networks, of information retrieval, and so on. As an alternative, randomized or stochastic processes have emerged as a promising new approach to these problems\cite{halko_martinsson_tropp,mahoney_drineas,avron_maymounkov_toledo}. Regardless of these developments within big data science, the field of physics has been grappling with these issues for decades, and have come up with a number of ideas for these types of problems. We believe a cross-fertilization of ideas from computer science, applied mathematics, and physics, will enable much more rapid growth in the ability to solve extreme data science problems.

 One physics-based example that fits into the extreme data paradigm is the nonequilibrium  quantum
many-body problem. This problem governs many different behaviors in physical systems.  Some examples include the following: (1) the current-voltage characteristic for transport of charge through a device made from strongly correlated materials; (2) the response functions for pump/probe experiments, where intense lasers are shone onto strongly correlated materials; (3) the response of a cold atom system following the quench of an interaction generated by a Feshbach resonance; (4) the response of materials to shock compression; (5) the evolution and thermalization of the quark-gluon plasma in heavy ion collisions; (6) particle production and the reheating/thermalization of the universe after inflation; (7) photo-ionization of atoms or molecules; and (8) the dynamical behavior of charged plasmas created by intense lasers.

It is well known that the Hilbert space for the many-body problem grows exponentially with the number of lattices sites in the problem. This growth is so rapid, that supercomputers can only handle problems that have a
few tens of lattice sites before the memory is completely exhausted. Such small systems often are too small to be able to reliably predict properties of $10^{23}$ particles. Fortunately, because of physical principles like those found in the renormalization group, many equilibrium many-body problems are governed by effective low-energy field
theories, where the high-energy degrees of freedom are frozen out, and do not change as the temperature is further
lowered. Using these ideas, the numerical renormalization group showed how one could numerically solve the
Kondo problem, which is one of the most important many-body problems in condensed matter physics. A large number of other techniques
exist for treating equilibrium many-body problems, and they often are governed by the ideas of the renormalization group,
via energy minimization methods where the quantum problem can be mapped onto an equivalent statistical problem
that can be solved using importance sampling methods like the Metropolis algorithm, or via variational
techniques chosen to minimize the loss of information like the density matrix renormalization group. 

In the case of nonequilibrium,
problems become more complicated, because energy minimization is often no longer a valid physical principle, and 
the driving of the system pushes it out of the low-energy effective field theory regime
(of course, reducing information loss is probably still a valid principle). At first glance, it then
seems hopeless to try to solve such problems, but significant progress has been made recently with a number of 
different algorithms, even if they all are currently limited by how far out in time they can evolve the system. In
most cases this time scale is too short to fully determine the long-time behavior. Nevertheless, it is highly likely that even a nonequilibrium system will only evolve through a very {\it small target subspace} of the infinite dataspace (especially if entanglement growth is bounded), and hence there is
a good chance that appropriate algorithms can be developed to describe the behavior of these systems.
In fact, there is a proof\cite{verstraete} that states that if the Hamiltonian involves local interactions (or interactions cut off after a short range), then the time-dependent evolution of the system can only expand into an exponentially small subspace of the full Hilbert space in polynomial time.  Techniques that can focus in on just this subspace will be the most efficient in simulating the nonequilibrium many-body problem.

When viewed from the computer-science perspective, extreme data science is based on four principals: (i)  encoding the active target subspace of the dataspace; (ii) evolving the target subspace forward in time; (iii) discarding nonphysical dimensions of the dataspace
where the target has negligible weight; and (iv) extrapolating (or forecasting) the behavior of the target space for
future times from the past history. Algorithms that solve the nonequilibrium quantum many-body problem often employ these
extreme data science principles in formulating their solution. For example, since one cannot use all of the basis vectors of the Hilbert space to represent a state vector, one must find an efficient method to encode the many-body wavefunction
which can describe the system with a small set of basis functions initially, and can evolve to remain efficient in describing the 
system as it moves forward in time (similar in spirit to compressive sensing). All algorithms formally deal with the evolution of the system in time, as
given by the quantum evolution operator 
\begin{eqnarray}
U(t,t_0)&=&\mathcal{T} \exp \left [ -\frac{i}{\hbar}\int_{t_0}^t dt' \mathcal{H}(t')\right ]\\
&=&\sum_{n=0}^\infty \left (-\frac{i}{\hbar}\right )^n\int_{t_0}^t dt_1\int_{t_0}^{t_1} dt_2
\int_{t_0}^{t_2}dt_3\cdots \int_{t_0}^{t_{n-1}} dt_n \mathcal{H}(t_1)\mathcal{H}(t_2)\cdots \mathcal{H}
(t_n)\nonumber
\label{eq: evolution}
\end{eqnarray}
which is a time-ordered exponential of the time-dependent Hamiltonian $\mathcal{H}(t)$, because this is how one ultimately determines the Green's function, or the evolution
of the wavefunction. Third, algorithms often include some form of truncation step, where the system is projected onto
a subsystem that includes the most important terms. Finally, since it is difficult to reach long times with most nonequilibrium
algorithms, one often attempts to extrapolate results either to try to stabilize the algorithm so it can run further in
time, or to try to forecast future behavior from the prior knowledge.

In the remainder of this review, we describe a wide range of different algorithms that have been used for the 
quantum many-body problem, describe their successes or failures for the nonequilibrium systems, and illustrate
a new attack motivated by extreme data science to create new algorithms that might have a higher chance for success.
We group the algorithms into those that simplify the many-body problem to solve it approximately, those that employ
stochastic methods and the challenges with generalizing them to the nonequilibrium case, algorithms based on 
renormalization group ideas, and algorithms that directly evaluate the evolution operator.  Note that while we have made every attempt to include a wide number of references to the original literature, this review is intended to be illustrative and not exhaustive, so many appropriate references have not been included.  We apologize to any authors for whom we did not reference their relevant work due to length restrictions.  In addition, we have not made an exhaustive review of big data science ideas as applied to the many-body problem.  For example, machine learning ideas have recently been employed to accelerate the solution of the problem\cite{millis_machine}. Instead, we concentrate on the new extreme data science aspect of the problem.

We emphasize that the techniques reviewed here focus on methods that work directly or indirectly in Hilbert space.
Recently, Gaussian phase space methods\cite{gaussian,gaussian2} have been proposed to evade the exponential
growth of the problem size in Hilbert space. They work by mapping the quantum evolution in real or imaginary time onto a set of stochastic differential equations with a drastically reduced dimensionality. Much excitement arose for this method early on, particularly with the anticipation that
phase-space methods could solve the fermionic sign problem\cite{sign}. But a careful analysis\cite{sde1,sde2} shows that
the nonvanishing of the distribution function evolved by stochastic differential equations at the boundaries,
introduces systematic errors such as spiking trajectories or rapid growth of the sampling error which currently 
prevent extracting reliable results for fermionic systems in the limits of strong interaction, large system sizes, or long times (the approach has been more successful for bosonic systems though\cite{drummond_boson}).  In any case, we focus the remainder of our
discussion on Hilbert-space-based methods.

\section{Algorithms that truncate the dataspace or simplify the many-body problem}

One strategy to take for an extreme data problem is to truncate or simplify it so that it results in a finite
dataspace that can fit onto the computer and be solved. Within the many-body problem, there are a few different 
approaches that fall into this category: (i) exact diagonalization of finite systems; (ii) configuration-interaction-based
methods; (iii) time-dependent density-functional theory; and (iv) the Gutzwiller variational method. We describe these approaches next.

Exact diagonalization is one of the first methods\cite{harris_lang} tried for many-body problems like the Hubbard model\cite{hubbard_model}. Early
work by the Falicov group\cite{falicov} or the work by Heilmann and Lieb\cite{heilmann} that investigated the properties of the benzene molecule, 
focused on determining energy eigenvalues and eigenstates after taking into account all of the symmetries of
the system. The technique has evolved significantly since then, and now has been applied to many nonequilibrium
problems where strongly correlated materials are driven by high intensity laser fields\cite{prelovsek,tohyama}. 

For equilibrium problems, one can handle systems with large Hilbert spaces because the Hamiltonian is sparse, and diagonalization methods like the Davidson\cite{davidson} or Lanczos\cite{lanczos} methods are iterative and require only the action of the Hamiltonian
on a state vector in the Hilbert space (in most of these calculations a number of vectors in the Hilbert space are stored in the computer memory).  This can often be calculated without even storing the Hamiltonian matrix,
when one uses the operator form and acts it on representative basis states. These techniques can only determine
eigenvalues (and associated eigenvectors) within a finite energy interval, and usually focus on the low-energy
states (in principle, long-run Lanczos methods that can compute all eigenvalues, or compute eigenvalues
within specified energy windows, do exist\cite{saad}, but they require too much storage to be efficient
methods for fully diagonalizing large sparse matrices). By using time evolution methods, such as evaluating the evolution operator via a Trotter expansion\cite{trotter}
(or more sophisticated methods like commutator-free exponential time expansions\cite{cfet}) or using the Crank-Nicolson algorithm\cite{crank_nicholson}
(which preserves unitarity through second order in the discretization time step),
one can evolve these finite systems forward in time and solve nonequilibrium problems. This has been looked at recently by a number of different groups\cite{prelovsek,tohyama}. 

The exact diagonalization approaches work with discrete eigenvalues and are accurate in equilibrium when the 
temperature is large relative to the average level spacing. For nonequilibrium, the time evolution is usually very 
accurate for short times, but then starts to deviate from the correct behavior when it displays recurrences, which 
occur for times such that the time multiplied by an energy difference is equal to a multiple of $2\pi$. Other issues
arise with finite-size effects if a disturbance moves a longer distance than half the distance to the edge of the 
finite system. While this approach can provide much insight about the nonequilibrium problem, it is unlikely
that it will ever be able to successfully describe generic systems in the long-time limit.

Another approach, which involves truncating the Hilbert space, is a configuration interaction-based approach, which is widely used in chemistry and atomic physics, but is not so widely used for
lattice systems. In this class of methods, the wavefunction is described by the sum of a finite number of Slater determinants, with the coefficients and the determinants adjusted to produce the best variational energy. These techniques are also used for time-dependent problems, primarily in atomic physics\cite{santra} where the wavefunction is represented as a time-dependent function with an
expansion basis that includes a description of the active excited electron and additional single-pair particle-hole
excitations or uses a so-called restricted active Hilbert space\cite{bonitz}. This creates a finite expansion for time-dependent coefficients of the wavefunction, which
are then solved by recasting the problem as a differential equation for the different coefficients and
solving those differential equations.  In practice, many tricks need to be implemented to solve these
problems accurately and efficiently. Within lattice problems, there have been some attempts to try
to adopt these approaches in nonequilibrium. One approach is called the multiconfiguration time-dependent Hartree impurity solver\cite{eckstein}, which employs a tensor-based structure for the
time-dependent coefficients of a wavefunction that is expanded in terms of a time-varying single-particle
basis set which is truncated to include just a small number of single-particle basis functions. While this approach shows some promise, much more work needs to be done to see if it can be an effective
impurity solver.

Another class of approaches is based on density functional theory and is called time-dependent
density functional theory (TD-DFT). Density functional theory is based on the Hohenberg-Kohn theorem\cite{hohenberg_kohn}
which states that the wavefunction is a unique functional of the electron density, and is implemented by
mapping the system onto a noninteracting problem via the Kohn-Sham framework\cite{kohn_sham}. TD-DFT has its origins in the Runge-Gross theorem\cite{runge_gross}, which applies to time-dependent
potentials. In practice, because density-functional theory cannot be solved exactly, it ends up being
an effective mean-field theory approach to the problem which is highly accurate for real materials
calculations in weakly correlated systems. The TD-DFT needs modification to handle the full
time dependence of the coupling to an electromagnetic field, and currently has not been applied
to too many problems in this category. Current implementations also use functionals that typically limit the applicability to finite closed quantum systems. It remains the most accurate approach for real systems calculations, but is known to fail when correlations become too strong, and hence will also not
be able to solve the most general nonequilibrium many-body problems. A recent review\cite{ulrich_yang}, provides a nice introduction to this topic. We will have more to say about
effective mean-field theory descriptions below.

The final method in this approximate approach is the Gutzwiller approach\cite{schiro,rigol}, which 
uses a variational form of the wavefunction for strong correlations, and projects out the double-occupancies in the wavefunction. This {\it ansatz} is then used as a constrained wavefunction to 
approximately solve the time-dependent Schr\"odinger equation via a minimization of the 
Schr\"odinger equation when taken as an expectation value with respect to the time-dependent
wavefunction. This approach has been applied primarily to quench problems, where the 
Hamiltonian during the evolution of the system is time independent, the initial state is just not
an eigenstate of this Hamiltonian, so it evolves in time. This technique is an uncontrolled
approximation, since it is not a formal expansion in powers of a small parameter, but it does
appear to produce physically reasonable results for the cases it has been applied to.

In all of these approaches, the many-body problem has been simplified in some fashion, allowing for the 
subsequent dynamics of the simplified problem to be computed exactly. These approaches produce
physically reasonable results, and often are quite accurate for short times, but they also
are difficult to gauge their accuracy unless they can be compared to more exact methods. In
general, they are not methods that can be easily extended to treat the problem 
with systematically improving approximations, so that it can eventually be solved in the
general case. However, they often are a good first approach to these problems.

\section{Algorithms that use stochastic methods}

We next turn to stochastic methods that solve the many-body problem via Monte Carlo techniques.
We begin by discussing how these algorithms work in equilibrium, before discussing issues that arise
for nonequilibrium algorithms. We also discuss only a few categories of these methods due to space
limitations. Our focus will be primarily with many-body approaches for the dynamical mean-field theory
technique, which provides an exact solution to the many-body problem in the limit of large
spatial dimensions. If one were to look at electrons moving on a lattice in real space, and were to
focus on what happens at a specific lattice site as a function of time, one would see electrons hopping onto
and off of the site in time. The dynamical mean-field theory approach maps this lattice problem onto
a single-site problem with a time-dependent field that represents the hopping onto and off of the
site.  By solving the impurity problem in a self-consistently determined time-dependent field that
makes the impurity Green's function identical to the local lattice Green's function, the problem is
solved exactly in the large dimensional limit (the reason why it is not exact in finite dimensions is that
the self-energy for the impurity problem has no momentum dependence, and hence is an approximate
solution in finite dimensions, but becomes exact in infinite dimensions because one can prove the
self-energy has no momentum dependence in that limit).

Impurity problems have been studied in physics for years. The single impurity Anderson model\cite{siam} is
the pardigmatic impurity problem, and it corresponds to a localized electron, which has a strong Coulomb
interaction with itself, embedded in a conduction electron host, that is noninteracting (in the strong-interaction limit, this model reduces to the so-called Kondo problem\cite{kondo}). These impurity problems are best represented by an action, which takes into account the time-dependence of the 
field that describes the hopping onto and off of the lattice site. Due to the interaction of the impurity electron with itself, there is a four-fermion operator in the action.  To illustrate how this representation works, we must first start with the definition of the contour-ordered Green's function on the lattice, which is defined
via
\begin{equation}
G_{ij\sigma}^c(t,t')=-i{\rm Tr}\mathcal{T}_c e^{-\beta\mathcal{H}(\bar t\rightarrow -\infty)}c_{i\sigma}^{}(t)c_{j\sigma}^\dagger(t')/\mathcal{Z}
\label{eq: g_contour}
\end{equation}
where the fermionic destruction operator on site $i$ with spin $\sigma$ $c_{i\sigma}^{}$ and the fermionic creation operator on site $j$ with spin $\sigma$ $c_{j\sigma}^\dagger$ are both written in the Heisenberg representation
with respect to the time-dependent Hamiltonian $\mathcal{H}(t)$, $\beta=1/T$ is the inverse
temperature of the initial equilibrium configuration of the system, and the partition function is
$\mathcal{Z}={\rm Tr}\exp[-\beta \mathcal{H}(\bar t\rightarrow -\infty)]$. The time-ordering operator orders times along the contour, which will be described next. 

The Heisenberg representation of an operator satisfies $c_{i\sigma}^{}(t)=U^\dagger(t,-\infty)c_{i\sigma}^{}U(t,-\infty)$. Substituting this into the definition of the Green's function and assuming $t>t'$ yields
\begin{equation}
G_{ij\sigma}^c(t,t')=-i{\rm Tr}e^{-\beta\mathcal{H}(\bar t\rightarrow -\infty)}U(-\infty,t)c_{i\sigma}^{}
U(t,t')c_{j\sigma}^\dagger U(t',-\infty)/\mathcal{Z},
\end{equation}
where we note that we used the fact that $U^\dagger(t,-\infty)=U^{-1}(t,-\infty)=U(-\infty,t)$ and
$U(t,t_1)U(t_1,t')=U(t,t')$. Now, starting from the earliest time as $\bar t=-\infty$, which corresponds to the right most part of the above expression, we evolve the system forward in time from $-\infty$ to $t'$, then we operate with a $c_{j\sigma}^\dagger$, evolve further to time $t$, operate with
$c_{i\sigma}^{}$ and then evolve backwards in time from $t$ to $-\infty$ before operating with the density matrix operator. This can be summarized graphically by evaluating the operator average on the
so-called Kadanoff-Baym-Keldysh contour (see Fig.~1), to which we have added an imaginary spur from $-\infty$ to $-\infty-i\beta$, and the density matrix term can be thought of as
\begin{equation}
e^{-\beta\mathcal{H}(\bar t\rightarrow -\infty)}=\exp \left [ -i\int_{-\infty}^{-\infty-i\beta}d\bar t \mathcal{H}(\bar t)\right ],
\end{equation}
so that the evolution operator extends along the entire contour. We originally thought of $t$ and $t'$ both being real times, but now that we have the evolution operator evolving on the contour, we can generalize the definition to allow the times $t$ and $t'$ to be any two times chosen on the contour itself, and the formula for the Green's function can be written compactly as
\begin{equation}
G_{ij\sigma}^c(t,t')=-i{\rm Tr}\mathcal{T}_c \exp\left [ -i\int_c d\bar t \mathcal{H}(\bar t)\right ] c_{i\sigma; t}^{} c_{j\sigma; t'}^\dagger /\mathcal{Z}
\end{equation}
where the contour now extends from $-\infty$  to $\infty$ along the upper real branch, from $\infty$
to $-\infty$ along the lower real branch, and from $-\infty$ to $-\infty-i\beta$ along the imaginary
spur.  The integral is over the full contour, and the subscript $t$ and $t'$ indicate where the operators 
act along the contour, with the time-ordering operator along the contour distributing the (time-ordered) exponential
factors in the right locations between the operators.

\begin{figure}[bt]
\centerline{\includegraphics[width=2.5in,clip]{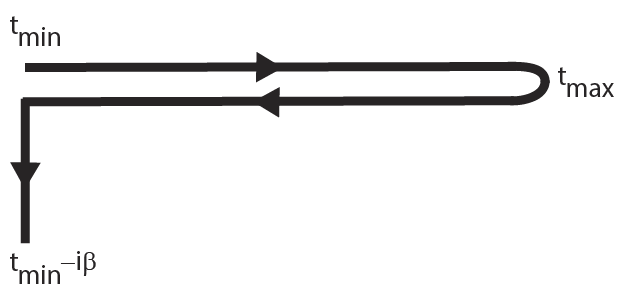}}
\vspace*{8pt}
\caption{The Kadanoff-Baym-Keldysh contour running from $t_{\rm min}$ to $t_{\rm max}$ along the real axis, and then back from $t_{\rm max}$ to $t_{\rm min}$ along the real axis, and finally along the negative imaginary axis a length $-i\beta$. We take the limit $t_{\rm min}\rightarrow -\infty$ and $t_{\rm max}\rightarrow\infty$. The system starts in equilibrium at a temperature $T=1/\beta$ at time $t_{\rm min}$ and then is driven out of equilibrium via a quench, or an applied field, or some other perturbation.}
\end{figure}

In dynamical mean-field theory, we will equate the local Green's function $(i=j$) to the impurity
Green's function.  The impurity Green's function depends on the dynamical mean field, which is 
often described by the symbol $\lambda_\sigma(t,t')$ with either time on the contour. Since the dynamical mean field determines the effect of the hopping on the lattice, we define the local part of the
Hamiltonian on the lattice via $\mathcal{H}_0^i$ such that the total Hamiltonian on the lattice
is equal to $\mathcal{H}(t)=\sum_i\mathcal{H}_0^i+\mathcal{K}$ where $\mathcal{K}$ is the kinetic energy term of the Hamiltonian. The local part of the Hamiltonian has the same functional form on each lattice site; the $i$ superscript simply denotes that we construct it from the operators $c_{i\sigma}^{}$ and $c_{i\sigma}^\dagger$. We then take the impurity Hamiltonian to be $\mathcal{H}_0$ constructed from the impurity creation and annihilation operators $c_\sigma^\dagger$ and $c_\sigma^{}$, respectively. All operators for the impurity problem are taken in the Heisenberg representation 
{\it with respect to} the impurity Hamiltonian. Then the impurity Green's function is written as
\begin{equation}
G_{\sigma;imp}^c(t,t')=-i{\rm Tr}\mathcal{T}_c e^{-\beta\mathcal{H}_0(\bar t\rightarrow -\infty)}
e^{\sum_\sigma\int_c d\bar t \int_c d\bar t' c^\dagger_\sigma(\bar t)\lambda_\sigma(\bar t,\bar t')c_\sigma(\bar t')} c_\sigma^{}(t)c_\sigma^\dagger(t')/\mathcal{Z}_{imp}
\end{equation}
where $\mathcal{Z}_{imp}={\rm Tr}\mathcal{T}_c\exp[-\beta\mathcal{H}_0(\bar t\rightarrow\-\infty)]
\exp[ \sum_\sigma\int_c d\bar t \int_c d\bar t' c^\dagger_\sigma(\bar t)\lambda_\sigma(\bar t,\bar t')c_\sigma(\bar t') ]$, which can be restricted to integrating only over the imaginary part of the contour
for the partition function because the integration over the real parts cancels.

The dynamical mean field $\lambda_\sigma$ must be determined self-consistently in the dynamical mean-field
theory algorithm (see Fig.~2).  The procedure is as follows: (1) make a guess for the local self-energy; (2) use Dyson's equation in momentum space on the lattice to determine the momentum-dependent Green's function from the self-energy; (3) sum the momentum dependent Green's function over all momenta to determine the local Green's function of the lattice; (4) use Dyson's equation for the local Green's function to extract the dynamical mean field; (5) solve the impurity problem in the dynamical mean field for the
impurity Green's function; (6) use Dyson's equation to extract the impurity self-energy from the
Green's function and the dynamical mean field; (7) use the impurity self-energy as the new guess for step (1) and iterate steps (2--7) until one reaches a fixed point and the Green's functions do not change from one iteration to the next.  All of the steps of this algorithm are straightforward to implement (even if we have not given the detailed equations) except for step (5) which is what requires a sophisticated impurity solver. The iterative approach is the same for equilibrium or nonequilibrium, it is just that in nonequilibrium one usually works entirely in a time-based formalism\cite{freericks_neq}. Note that if one is solving an equilibrium problem, one usually restricts the contour to
just the imaginary spur, and then uses an analytic continuation method like maximum entropy\cite{maximum_entropy} to
numerically analytically continue the results from the imaginary axis to the real axis.  This simplifies the
impurity problem solver, but requires an ill-posed and uncontrolled numerical analytic continuation procedure. This is often viewed as
a superior way of performing the calculation than simulating the problem on the full contour, for
reasons that will be described below, which are related to the fermion sign (or phase)  problem.

\begin{figure}[bt]
\centerline{\includegraphics[width=3.65in,clip]{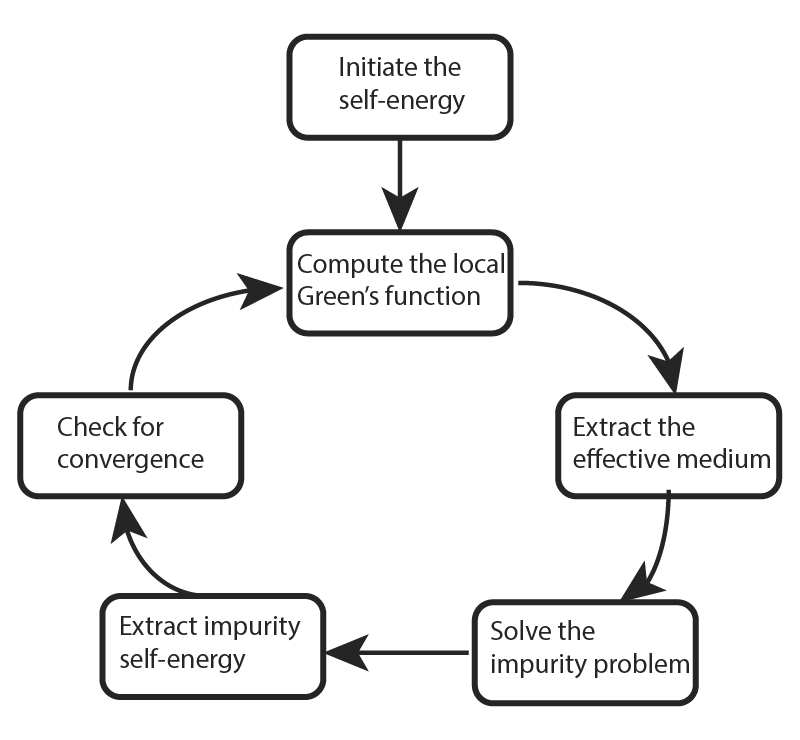}}
\vspace*{8pt}
\caption{Schematic of the iterative loop for dynamical mean-field theory. We start with a guess for the self-energy and then use Dyson's equation on the lattice to calculate the local Green's function (by summing over all momenta). Dyson's equation is used on the local Green's function to extract the effective medium (which depends on the dynamical mean field). The impurity problem is then solved with this effective medium (or $\lambda$ field). Dyson's equation for the impurity is used to extract the self-energy and one checks to see if the iterative loop has converged.  If it has, the computation ends.  If not, the new self-energy is used in the loop and the process continues. }
\end{figure}

We next focus on two different algorithms that solve the impurity problem on the imaginary spur of the
contour---the Hirsch-Fye algorithm\cite{hirsch_fye} and the continuous-time Monte Carlo algorithm\cite{lichtenstein,werner}. To begin, we show how one evaluates the partition function
on the imaginary-time spur of the contour within the Hirsch-Fye algorithm
\begin{equation}
\mathcal{Z}_{imp}={\rm Tr}\mathcal{T}_\tau e^{-\beta\mathcal{H}_0}e^{-\sum_\sigma\int_0^\beta d\tau
\int_0^\beta d\tau'c^\dagger_\sigma(\tau)\lambda_\sigma(\tau,\tau')c_\sigma^{}(\tau)}.
\end{equation}
The imaginary time axis is discretized into $N$ time steps of length $\Delta\tau=\beta/N$, with $t_j=-\infty-i\Delta \tau j$ for $0\le j\le N$, which we write as $-\infty-i\tau_j$. Then we have
\begin{equation}
\mathcal{Z}_{imp}={\rm Tr}\mathcal{T}_\tau  e^{-\Delta\tau \sum_{j=0}^N \mathcal{H}_0^{sing}(\tau_j)}e^{-\Delta\tau\sum_{j=0}^N
\mathcal{V}(\tau_j)}e^{-(\Delta\tau)^2\sum_\sigma\sum_{i,j=0}^Nc_\sigma^\dagger(\tau_i)\lambda_\sigma(
\tau_i,\tau_j)c_\sigma^{}(\tau_j)},
\end{equation}
where $\mathcal{H}_0^{sing}$ is the quadratic (single-particle)  part of the impurity Hamiltonian (which we will take to be
$\mathcal{H}_0^{sing}(\tau)=-\mu\sum_\sigma c^\dagger_\sigma c^{}_\sigma$ for concreteness) and
$\mathcal{V}=\mathcal{H}_0-\mathcal{H}_0^{sing}$ is the interacting part of the Hamiltonian
[which we take to be $\mathcal{V}(\tau)=U(c^\dagger_\uparrow c^{}_\uparrow-1/2)(c^\dagger_\downarrow c^{}_\downarrow-1/2)$ for the Hubbard model, for concreteness]. Note that we have included an artificial time dependence to the impurity Hamiltonian even though it is a constant, independent of time, which is notationally convenient. The challenge with evaluating this partition function comes from the fact that the interaction $\mathcal{V}$ is not a quadratic function of the fermionic operators. It can be made quadratic by introducing an auxiliary field via the Hirsch-Hubbard-Stratanovich transformation\cite{hirsch,hubbard,stratanovich}
\begin{equation}
e^{-\mathcal{V}(\tau_j)}=e^{-\Delta\tau U(c^\dagger_\uparrow c^{}_\uparrow-1/2)(c^\dagger_\downarrow c^{}_\downarrow-1/2)}=\frac{e^{-\Delta\tau U/4}}{2}\sum_{s_j=\pm 1}e^{\alpha s_j(c^\dagger_\uparrow c^{}_\uparrow-c^\dagger_\downarrow c^{}_\downarrow)},
\end{equation}
where $\cosh\alpha=\exp[\Delta\tau U/2]$. This introduces a set of Ising variables at each $\tau_j$
time step which converts the quartic interaction term into a quadratic term. One can now evaluate the
partition function using standard techniques that convert into a Grassmann path integral and then
evaluate the trace with coherent states to yield
\begin{equation}
\mathcal{Z}_{imp}=\sum_{\{s_i\}}\det \left ( G_{\uparrow\{s_i\}}^{-1}\right )
\det \left ( G_{\downarrow\{s_i\}}^{-1}\right )
\end{equation}
where 
\begin{equation}
G^{-1}_{\sigma\{s_i\}}(j,k)=\delta_{jk}-\delta_{j-1k}[1+\Delta\tau\mu+\alpha s_j\sigma]
-(\Delta \tau)^2\lambda_\sigma(\tau_j,\tau_k).
\end{equation}
So, the evaluation of this sum over all of the Ising spins will give the partition function. It turns out that averaging the Green's function $G$ using the weights for the partition function, will yield the impurity
Green's function. The Monte Carlo method starts with a configuration for the Ising spins, and proposes the change of the Ising spin at a particular time slice. One calculates the ratio of the determinants for the previous $G$ and the current $G$, using a Metropolis algorithm\cite{metropolis} to decide whether or 
not to accept the update move. If it is accepted, the Green's function and determinants need to be updated. 
One of the hallmarks of the Hirsch-Fye algorithm is that the updating of the Green's function and of the
determinant is very fast, because only a few matrix elements are changed from the previous configuration matrix.  Hence, because one need not
recalculate the full determinant from scratch, nor recalculate the new contribution to Green's function (for the new Ising configuration) from scratch, the algorithm can work fast to sum over the Ising fields via
importance sampling. 

Since it is a statistical algorithm, it must deal with probabilities that satisfy a detailed balance procedure,
so that every configuration is equally likely to be reached by the algorithm updating procedure. This
requires (among other things) that the products of the determinants be nonnegative.  Unfortunately, although they are
manifestly real, they often change sign as the system is updated. One uses the absolute value for the
weights, but one then must keep track of the average sign during the updating procedure and divide all
statistical averages by the average sign.  When this average sign becomes too small in magnitude, the 
algorithm breaks down.  This is the origin of the famous sign problem in quantum Monte Carlo, and
it limits how low in temperature one can run the simulation because the sign problem becomes worse as the
temperature is lowered. In addition, since one is using a fixed grid in imaginary time, it turns out one
needs to make $\Delta \tau$ small at low temperature as well, because the Green's function has a large slope
near $\tau=0$ and $\tau=\beta$ and one needs to determine it accurately in this region. This also limits how
low in temperature one can proceed with the Hirsch-Fye algorithm, since a small $\Delta \tau$ and a large
$\beta$ implies the size of the matrices is large.

We next discuss the continuous time quantum Monte Carlo algorithm in equilibrium and based on a 
weak-coupling expansion\cite{lichtenstein}. The starting point for the continuous-time algorithm is 
the formula for the impurity partition function, written in the interaction representation with respect to
$\mathcal{H}_0^{sing}$:
\begin{equation}
\mathcal{Z}_{imp}={\rm Tr}\mathcal{T}_\tau e^{-\beta\mathcal{H}_0^{sing}}
e^{-\int_0^\beta d\tau \mathcal{V}_I(\tau)}e^{-\sum_\sigma \int_0^\beta d\tau \int_0^\beta d\tau'
c^\dagger_\sigma(\tau)\lambda_\sigma(\tau,\tau')c_\sigma^{}(\tau')}
\end{equation}
where the time ordering properly orders both of the time-dependent exponential factors. Here, the interaction
picture interaction satisfies $\mathcal{V}_I(\tau)=\exp[\tau\mathcal{H}_0^{sing}]\mathcal{V}\exp[-\tau
\mathcal{H}_0^{sing}]$ and similarly for the fermionic creation and annihilation operators. We assume the concrete forms for the single-particle piece of the
impurity Hamiltonian and for the interaction piece as we did before, to describe the Hubbard model. The next step in the derivation  is to expand the exponential of the interaction in a power series
\begin{eqnarray}
\mathcal{Z}_{imp}&=&\sum_{k=0}^\infty \frac{(-U)^k}{k!}\int_0^\beta d\tau_1\ldots \int_0^\beta d\tau_k {\rm Tr} \mathcal{T}_\tau  e^{-\beta\mathcal{H}_0^{sing}}e^{-\sum_\sigma \int_0^\beta d\tau \int_0^\beta d\tau'
c^\dagger_\sigma(\tau)\lambda_\sigma(\tau,\tau')c_\sigma^{}(\tau')}\nonumber\\
&\times&c^\dagger_\uparrow(\tau_1)c^{}_\uparrow(\tau_1)\ldots c^\dagger_\uparrow(\tau_k)c^{}_\uparrow(\tau_k)
c^\dagger_\downarrow(\tau_1)c^{}_\downarrow(\tau_1)\ldots c^\dagger_\downarrow(\tau_k)c^{}_\downarrow(\tau_k).
\end{eqnarray}
Because this average over the product of the fermionic creation and annihilation operators
is with respect to a quadratic action (given by $\mathcal{H}_0^{sing}$ and the $\lambda$-field term),
one can evaluate the average via Wick's theorem, which expresses the result in terms of 
products of the Green's function corresponding to the quadratic action. In total, there are 
$k!$ different ways to form these products, but they can all be summarized via the determinant of an
appropriate matrix. Note that here we have
\begin{equation}
G_{0\sigma}^{-1}(\tau,\tau')=(-\partial_\tau+\mu)\delta(\tau-\tau')-\lambda_\sigma(\tau,\tau').
\end{equation}
We define the determinant $D_\sigma(k)$ via
\begin{equation}
D_\sigma(k)={\rm Det}\left ( 
\begin{array}{cccc}
G_{0\sigma}(\tau_1,\tau_1)&G_{0\sigma}(\tau_1,\tau_2)&\cdots&G_{0\sigma}(\tau_1,\tau_k)\\
G_{0\sigma}(\tau_2,\tau_1)&G_{0\sigma}(\tau_2,\tau_2)&\cdots&G_{0\sigma}(\tau_2,\tau_k)\\
G_{0\sigma}(\tau_3,\tau_1)&G_{0\sigma}(\tau_3,\tau_2)&\cdots&G_{0\sigma}(\tau_3,\tau_k)\\
\cdot&\cdot&\cdot&\cdot\\
\cdot&\cdot&\cdot&\cdot\\
\cdot&\cdot&\cdot&\cdot\\
G_{0\sigma}(\tau_k,\tau_1)&G_{0\sigma}(\tau_k,\tau_2)&\cdots&G_{0\sigma}(\tau_k,\tau_k)\\
\end{array}
\right )
\end{equation}
with the time arguments given by the $k$ different time values that will be taken in the $k$th multiple
integral (see Fig.~3 for an example diagram). The stochastic aspect of the problem is now clear. One uses Monte Carlo methods
to sample the different $\tau_i$ values for each multidimensional integral, and since the error
decreases as the number of random points chosen in the integrand independent of dimension,
one can, in principle, evaluate each order of the perturbation theory expansion with an equivalent amount of computer
time.  This, of course, is not the case since the integrand involves the determinants, and those 
determinants are more difficult to evaluate for larger $k$.  In an actual implementation of the algorithm,
the Monte Carlo step allows for the system to increase or decrease $k$ by one unit (adding a vertex
or removing a vertex), and the acceptance criteria for the Monte Carlo step comes from evaluating the
ratio of the determinants for the two sets of diagrams evaluated. Fast updating methods are
used to calculate the new determinants from the old ones without having to diagonalize the full matrix,
which is what makes the algorithm efficient. The algorithm will actually determine the most important orders
of the perturbation theory via the importance sampling. 

\begin{figure}[bt]
\centerline{\includegraphics[width=3.65in,clip]{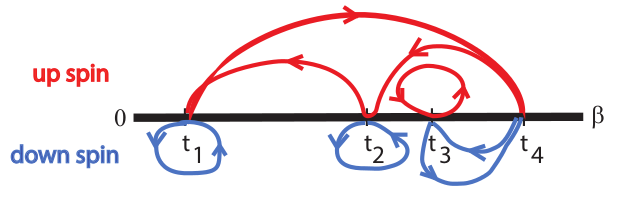}}
\vspace*{8pt}
\caption{(Color online) Example of one diagram in imaginary time for the equilibrium implementation of the
continuous time algorithm from weak coupling. The narrow lines are Green's functions connecting different time points (red for spin up and blue for spin down). The Ising variable associated with the auxiliary field at the interaction vertex has been suppressed. }
\end{figure}

Once again, there is a sign problem that occurs if the determinant changes sign, which often happens. To
handle this, the interaction is modified from the conventional Hubbard interaction to
\begin{equation}
U\left (c_\uparrow^\dagger c_\uparrow^{}-\frac{1}{2}\right )\left (c_\downarrow^\dagger c_\downarrow^{}-\frac{1}{2}\right )=\frac{U}{2}\sum_{s=\pm}\left (c_\uparrow^\dagger c_\uparrow^{}-\frac{1}{2}+s\alpha\right )\left (c_\downarrow^\dagger c_\downarrow^{}-\frac{1}{2}-s\alpha\right ),
\end{equation}
where $\alpha\ge 1/2$ seems to fix the sign problem for the equilibrium calculations.  Of course, this does require the additional summation over Ising variables at each time step for a given diagram (and a rederivation of the above formalism using the new interaction).  The summation is also
done via stochastic importance sampling.

Calculating the Green's function is a bit more involved for the continuous-time algorithm. The formula for the Green's function includes two more creation and annihilation operators than the 
partition function.  This requires one more Wick contraction for each diagram of ``order'' $k$ than
used in the partition-function calculation.  Otherwise, the technique is identical.  By manipulating the
Dyson equation, one can show that the term that is actually summed in the algorithm is a reducible
self-energy, which can readily be employed to determine the full Green's function.

One might ask about the relationship between these two methods. It turns out that they have
a simple relationship\cite{mikelsons}. If one were to restrict the continuous time algorithm to include only diagrams of order $N=\beta/\Delta\tau$ and allow the vertices to be restricted to the time
slices used in the Hirsch-Fye algorithm, then the two methods calculate the same set of diagrams. By extrapolating to the $N\rightarrow\infty$ limit, one can then show that the extrapolated
results for these two methods will be identical, because the continuous time method will require 
diagrams of effectively one order only, due to the central limit theorem. It is also clear, then, that
the continuous time algorithm should be superior to the Hirsch-Fye algorithm for truncated
calculations, because it effectively samples a wider range of diagrammatic terms. The continuous-time
approach has now become the standard technique used by most dynamical mean-field theory calculations in equilibrium.

There is a second form for the continuous-time algorithm, which performs the expansion of the perturbation series from a strong-coupling approach called the hybridization expansion\cite{werner}. We do not have the space here to describe the derivation of this approach.  While similar to the weak-coupling technique, the derivation is more complicated because 
the atomic Hamiltonian is not quadratic, and hence one cannot use Wick's theorem, but instead one has to employ some more complex techniques. 

Since the continuous-time algorithm is a diagrammatic method, which sums diagrams to high order, one might wonder whether or not there might be a better way to organize the diagrams, so that an infinite class of diagrams can be summed initially, and the Monte Carlo is used to sum the remaining ones.  This idea is called the bold quantum Monte Carlo method\cite{prokofiev,gull1}. The idea behind the bold method, is that if one can identify the dominant terms in the summation that provide the physical behavior of the system, then summing them up front can greatly reduce the order that one needs to extend the quantum Monte Carlo summation. There are additional challenges to employing the bold technique, as one now needs to identify and remove terms in the diagrammatic summation that already appear in the infinite summation of the starting point. This can make the diagram selection and the Monte Carlo integration more complicated, but this is usually more than compensated by requiring a much lower order in terms of the number of diagrams to keep in the expansion. Usually the latter is more important than the former, especially in cases where there is a sign problem to deal with.

Finally, we need to discuss how one generalizes these approaches to nonequilibrium. In principle,
the procedure is straightforward, where we simply extend the contour along the real axis and
proceed as before.  The main problem with this is that the matrix elements for the Green's functions
now become manifestly complex, and the weight factors also become complex. One can once again use the modulus of the complex numbers for the weights, but then one needs to sum the average
complex phase of the summations and divide by the average phase.  The average phase rapidly goes to zero as the real branch of the contour increases in size, and this poses the most severe limitation on these methods.  The first attempt in this area was to generalize the Hirsch-Fye algorithm along these lines\cite{rabani}. The continuous-time algorithm has been tried\cite{schiro2,eckstein_werner}, and it can only be stabilized for quite short times, where only specialized problems can be studied that have their transient evolution take
place over a short period of time. The bold technique has also been applied\cite{gull2} and
shows promise, being able to be evolved about twice as long as the conventional algorithm. But,
generically, more stable algorithms that can evolve further in time are needed within this class
of impurity solvers. It is possible that making the algorithm manifestly unitary can help with the phase problem.

The weak convergence or divergence of a diagrammatic series is usually handled by techniques like Borel, Ces\'aro-Riesz, or Lindel\"of resummation\cite{svistunov}. A more general approach to doing this is offered by  resurgence theory\cite{ecalle}, which expresses $f(x)$ as a triple power series (called a trans-series) in $x$, $\exp[-1/x]$, and $\ln[x]$, employs analytic properties in the vicinity of all poles and branch cuts to determine the relevant coefficients of the expansion, and automatically includes nonperturbative effects.  It is possible that employing this method within a bold-like quantum Monte Carlo approach could allow calculations to proceed to even longer times\cite{lebowitz}. We discuss these techniques in more detail below when we examine extrapolation of nonequilibrium results to longer times.

There also have been some recent proposals for extensions of variational quantum Monte Carlo approaches
to nonequilibrium. So far, this has been done only for bosonic systems\cite{italians1,italians2}, but there is promise that one might be able to extend these ideas to fermionic systems, and at the very least, the bosonic methods might benefit from being re-examined along the extreme data science perspective.

\section{Algorithms based on renormalization group ideas}

The next class of algorithms we discuss are those related to the renormalization group of Wilson\cite{wilson}.  The original numerical renormalization group was designed to treat
the low-energy scale that develops in a system of noninteracting conduction electrons (the bath electrons) that hybridize with an impurity atom that has strong Coulomb interactions for electrons on the impurity (the impurity electrons).
It was well known that this model had complex behavior with a low energy scale that develops
in the system at low temperature and that the low-energy model displays universal behavior. Numerically calculating such results turned out to be problematic. Wilson's idea was to discretize the frequency space into a logarithmic energy grid, and replace the continuum of states within each
energy interval by one discrete degree of freedom. As one moves in energy closer to the Fermi
energy, the energy intervals become smaller and smaller.  Hence, the Hamiltonian for the single impurity Anderson model is described
by
\begin{equation}
\mathcal{H}_{SIAM}=E_f\sum_\sigma f^\dagger_\sigma f^{}_\sigma+Uf^\dagger_\uparrow f^{}_\uparrow f^\dagger_\downarrow f^{}_\downarrow+\sum_{\alpha  =0}^N
\sum_\sigma V_\alpha \left ( f^\dagger_\sigma c^{}_{\alpha\sigma}+c^\dagger_{\alpha\sigma}f^{}_\sigma \right )+\sum_{\alpha =0}^N\sum_\sigma\omega_\alpha
c^\dagger_{\alpha\sigma} c^{}_{\alpha\sigma}
\end{equation}
where $0\le\alpha\le N$ labels the bath states that have energy which we choose for illustration to satisfy $\omega_\alpha=1/\Lambda^\alpha$ (the parameter $\Lambda$ is commonly chosen between 1.5
and 3), $V_\alpha$ is the hybridization between the impurity and the $\alpha$th bath state,
$U$ is the Coulomb interaction, and $E_f$ is the energy of the impurity level. In this form, 
the Hamiltonian is written as an impurity coupled to each bath site, which is schematically illustrated in Fig.~4(a).  One can integrate out the conduction electrons, which produces an impurity problem interacting with a time-dependent field
which is our old $\lambda$ field, but written in real frequency instead of time. One finds
\begin{equation}
\lambda_\sigma(\omega)=\sum_{\alpha =0}^N \frac{|V_\alpha|^2}{\omega-\omega_\alpha+i0^+}.
\end{equation}
The negative of the imaginary part of this field is called the hybridization function $\Delta_\sigma(\omega)=-{\rm Im}\lambda_\sigma(\omega)$. Given a dynamical mean-field
$\lambda_\sigma$ in frequency space, one can immediately construct the hybridization and energy for each bath state that approximates the imaginary part of the dynamical mean-field as a sum over a set of delta functions. The procedure is as follows: for the $\alpha$th energy interval, which runs from $\omega_\alpha^{min}$ to $\omega_\alpha^{max}$, the frequency $\omega_\alpha$ is chosen to
be the weighted average of the hybridization function over the interval
\begin{equation}
\omega_\alpha=\frac{1}{\omega_\alpha^{max}-\omega_\alpha^{min}}\int_{\omega_\alpha^{min}}^{\omega_\alpha^{max}}d\omega \omega\Delta(\omega)
\end{equation}
and the weight of the delta function that describes the effective discrete degree of freedom yields the hybridization
\begin{equation}
V_\alpha=\sqrt{\frac{1}{\omega_\alpha^{max}-\omega_\alpha^{min}}\int_{\omega_\alpha^{min}}^{\omega_\alpha^{max}}d\omega \Delta(\omega)},
\end{equation}
which is always real, because the hybridization function is always nonnegative [see Fig.~4(b) for a schematic of the process].

\begin{figure}[bt]
\centerline{\includegraphics[width=1.9in,clip]{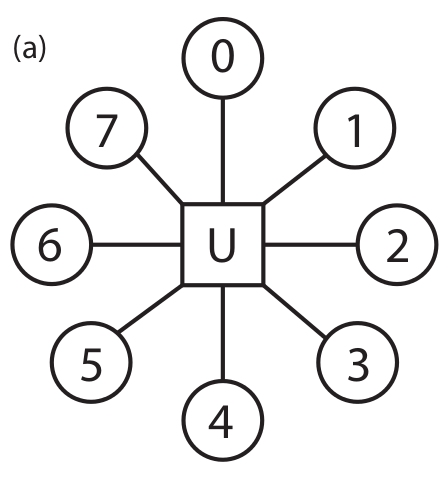}\includegraphics[width=2.6in,clip]{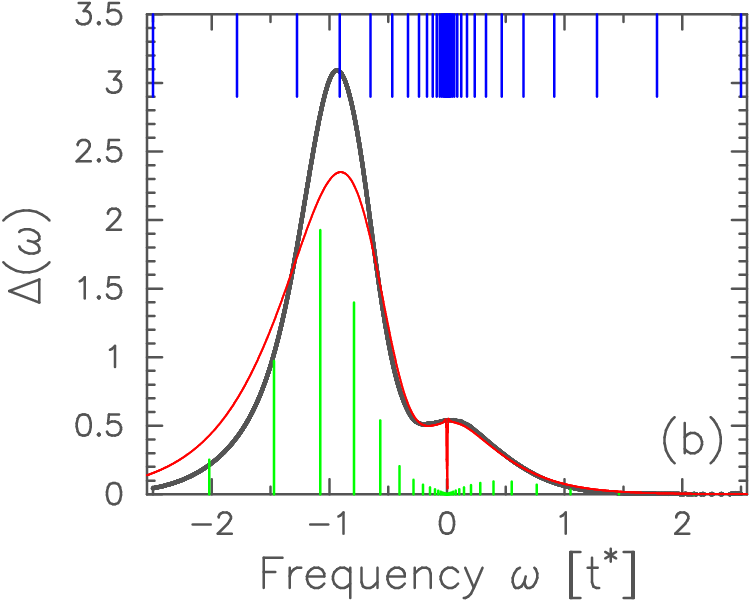}}
\vspace*{8pt}
\caption{(Color online) In this schematic of the single impurity Anderson model, we take eight bath states indicated by the circles which couple to the central impurity site indicated by the square, where the interaction takes place [panel (a)]. In addition, we show the traditional logarithmic grid in frequency space (upper blue) and how one determines the weight and frequency associated with each bath site (lower green) for the dynamical mean field of a Hubbard model with strong interaction and a filling of 0.9 [panel (b)]. 
The black curve is the original hybridization $\Delta(\omega)=-{\rm Im}\lambda(\omega)$ and the red curve is the approximation using $\Lambda=1.4$ and 25 points. The red curve is a reconstruction of the dynamical mean field by using a logarithmic broadening of the delta functions. Note how it fits poorly for large frequency but is much more accurate near $\omega=0$. Note also that there is a large deviation at the smallest frequencies because there are no discretized states there.  Increasing the number of states would fix that disagreement.}
\end{figure}

Wilson's next step is to construct the Wilson chain.  This is done by starting with the impurity Hamiltonian written in terms of the $f$ operators only, and then constructing the linear combination of bath states that directly couple to the impurity
\begin{equation}
a_{1\sigma}=\sum_{\alpha =0}^N V_\alpha c_{\alpha\sigma}^{}/\sqrt{\sum_{\alpha' =0}^N |V_{\alpha'}|^2}
\end{equation}
where the constant term in the denominator is chosen to guarantee that the $a$ operator satisfies the canonical fermionic anticommutation relations.  We now want to define a new set of chain
fermions (with $a_{0\sigma}=f_\sigma$) and construct the Wilson chain Hamiltonian (see Fig.~5), which takes the form
\begin{eqnarray}
\mathcal{H}_{SIAM}^{chain}&=&E_f\sum_\sigma a^\dagger_{0\sigma} a^{}_{0\sigma}+Ua^\dagger_{0\uparrow} a^{}_{0\uparrow} a^\dagger_{0\downarrow} a^{}_{0\downarrow}
+\sum_{i=1}^\infty \sum_\sigma t_i \left (a^\dagger_{i-1\sigma}a^{}_{i\sigma}+
a^\dagger_{i\sigma}a^{}_{i-1\sigma}\right ) \\
&+&\sum_{i=1}^\infty\sum_\sigma \epsilon_i a^\dagger_{i\sigma}
a^{}_{i\sigma}\nonumber
\end{eqnarray}
where each term in the Wilson chain is determined recursively by forcing the original single-impurity
Anderson model into the above form. The Coulomb interaction $U$ appears only on the zeroth lattice site, and the hopping matrix elements decay exponentially as one moves down the chain. One can
view this transformation as a tridiagonalization of the SIAM Hamiltonian in the same way
that one writes the Lanczos algorithm.

\begin{figure}[bt]
\centerline{\includegraphics[width=3.65in,clip]{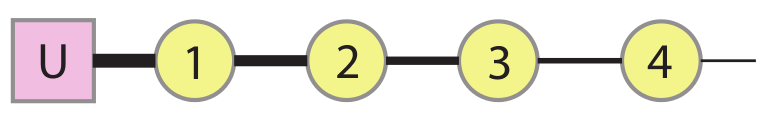}}
\vspace*{8pt}
\caption{ (Color online) The Wilson chain is constructed similar to a Lanczos procedure, where the impurity is coupled to a linear combination of bath sites, which we denote as the first chain site.  The chain site is then coupled in turn to new chain sites, with the coupling between chains decreasing down the chain (indicated by  the shrinking widths of the lines denoting the hopping).}
\end{figure}

Wilson's numerical renormalization group works with the problem in the Wilson chain form. One starts
with the impurity site and then adds one chain site and another, and so on, until the size of the matrix is just bigger than a predetermined size $m$ (typically $m$ ranges from 2000 to 5000) which is going to determine the number of kept states. One then diagonalizes the full Hamiltonian, and truncates the basis, keeping the $m$
lowest energy eigenvalues. The process is then iterated further, with another chain site added, the Hamiltonian constructed for the enlarged chain, then diagonalized, and finally truncated to the low-energy space. As this process is iterated, eventually it stops at a fixed point, where the system
is described by the renormalized Fermi liquid. By keeping account of the states discarded at each step of the iteration, one has a complete set of states for the problem that describes the high-energy degrees of freedom (which become frozen in at high temperature) and the low-energy degrees of freedom, which correspond to the active $m$ states on the current Wilson chain. Since one has 
both eigenvalues and eigenvectors, one can compute many different quantities of interest,
like the single-particle Green's function (via the Lehmann representation), the self-energy,
magnetic and charge susceptibilities, etc. Spectral properties are also recorded as a series of
delta functions which need to be broadened to obtain a spectral function. The broadening scheme needs to be chosen for a given problem, but usually one chooses a logarithmic scaled Lorentzian or Gaussian.  This choice does well at low frequencies, but does not properly produce the bandwidth of the system. 
 Since one can input the hybridization function, and output the self-energy, this method can be used to solve dynamical mean-field theory, although care needs to be taken in how this is done to ensure the iterated equations will converge.

One subtle point to keep in mind is that the effective temperature of the system is determined by the energy scale of the chain where the calculation ends. Hence, to get to the zero-temperature limit, one performs the calculation on a chain that is long enough that the results stop changing. Higher temperatures are found on smaller chains, and it should become clear that this technique fails if the temperature is too high, because there are not enough degrees of freedom to properly describe the behavior (in fact, for dynamical mean-field theory, it also fails at low temperature, as the imaginary part of the self-energy often has the wrong sign for frequencies near the Fermi energy at low temperature, below the low-energy coherence scale).

While the numerical renormalization group approach has been extremely successful in solving
some of the paradigmatic many-body problems, it turned out to be limited in scope, in that
similar ideas applied to strongly correlated lattice problems failed to work effectively. The
reason for this was twofold: (i) it isn't necessarily true that the low-energy eigenfunctions for a smaller system are the most important ones to keep for finding the low-energy eigenfunctions of a larger system and (ii) the choice for keeping the lowest energy eigenvalues turns out to significantly lose information about the quantum state as the system size grows. In the 1990's,
White developed an alternative formulation to the renormalization group, which he
called density matrix renormalization group, that solved both of these problems and allowed
a wide range of one-dimensional strongly correlated problems to be solved\cite{white}. A quantum information analysis of this algorithm shows that it works well as long as the entanglement entropy does not grow too rapidly.  This usually holds in one-dimensional systems, and rarely does for
higher dimensional ones. We describe the density matrix renormalization group algorithm next.

\begin{figure}[bt]
\centerline{\includegraphics[width=3.65in,clip]{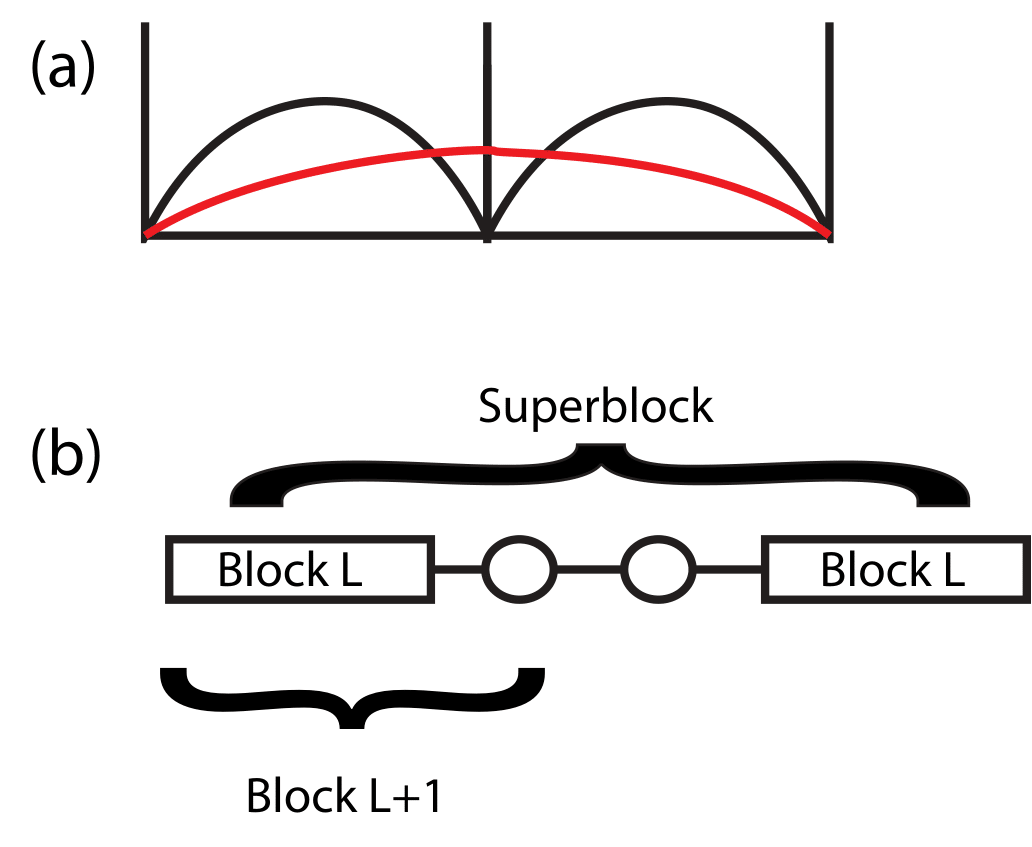}}
\vspace*{8pt}
\caption{(Color online) (a) Particle in a box problem which motivated the density matrix renormalization group approach. The ground state wavefunction in the smaller boxes (black curves)  do not allow for a simple way to construct the ground state wavefunction of the larger box  (red curve), although they do have some weight in that construction. (b) Schematic of the original block of size $L$, the additional site, the final block of size $L+1$ and the reflection of the system to construct a superblock.}
\end{figure}

The motivation for the density matrix renormalization group problem was the challenge of finding a
way to apply the renormalization group ideas to the case of expanding the size of a one-dimensional 
quantum system and maintain an appropriate basis that projects primarily onto the low-energy
states of the larger system. Consider a particle in a box of length $L$ in the ground state. We want to 
double the size of the system to $2L$ and find the states in the low-energy manifold.  One might think that they should consist of the states that have low energy in the smaller box, but a simple calculation shows that smaller box ground state projects onto states of all energies in the larger box, and hence
choosing states to keep by minimizing the energy is not the correct thing to do [see Fig.~6(a)]. In fact, the entire renormalization group procedure needs to be modified. White chose to increase the system of size $L$ to one of size $L+1$ through a complicated procedure involving a superblock of size $2L+2$. One starts with the block of size $L$, reflects it to another block of size $L$ and adds two sites in the middle. If we let $\alpha$ and $\beta$  denote the states in the left and right blocks of size $L+1$, then the wavefunction of the superblock is given by the matrix-product state expression
\begin{equation}
|\psi\rangle_{2L+2}=\sum_{\alpha\beta }\psi_{\alpha\beta }|\alpha\rangle \otimes |\beta\rangle
\end{equation}
where $\psi_{\alpha\beta}$ are the expansion coefficients (numbers) and $|\alpha\rangle$ and
$|\beta\rangle$ are the states that form the basis in the left and right block [see Fig.~6(b)].
Using a singular value decomposition, one can transform this matrix into its diagonal form
\begin{equation}
|\psi\rangle_{2L+2}=\sum_{\bar\alpha}\lambda_{\bar\alpha}|\bar\alpha\rangle_{\rm left}\otimes
|\bar\alpha\rangle_{\rm right}
\end{equation}
where the barred symbols are the rotated basis, and the same states are coupled together in the
direct products for the left and right subblocks. The reduced density matrix for the left block is then defined to be
\begin{equation}
\rho_{left}=\sum_{\bar\alpha_{\rm right}}|\psi\rangle\langle\psi |
=\sum_{\bar\alpha}\lambda^2_{\bar\alpha}|\bar\alpha\rangle_{\rm left} \,_{\rm left}\langle \bar\alpha |.
\end{equation}
The density-matrix renormalization group algorithm says that one takes the $m$ largest
eigenvectors of the reduced density matrix as the truncated basis for the left block of size $L+1$.
So the procedure is to start with a block of size $L$ that has already been written in the effective basis according to the maximal eigenvalues of the reduced density matrix. Construct a superblock of size 
$2L+2$ and diagonalize the Hamiltonian to give the ground state via a Davidson or Lanczos procedure,
take the matrix that expresses the ground state wavefunction in the original basis and diagonalize via
a singular value decomposition and transform the matrix to this new basis. Construct the reduced density
matrix by tracing over the right degrees of freedom and pick the $m$ states with the largest eigenvalues for the basis states of the $L+1$ block.  Proceeding in this fashion, one can keep growing the system until the energy stops changing. This then gives the results for the ground state of the equilibrium and infinite-sized system. One can examine multiple low energy states either by separately calculating for the different individual states, or by averaging the density matrix over the different eigenstates one is computing.

There also is a finite-sized system algorithm that also creates superblocks, but the size of the system is fixed at $2L$ say, so one increases the left block while decreasing the right and vice versa, zipping through the system to the left and then to the right. This approach is usually more accurate for determining the energy, especially if one uses finite size scaling to the large system limit, than the infinite system algorithm. 

The density-matrix renormalization group algorithm is a variational algorithm, and the choice of using the reduced density matrix rather than the energy eigenvalues in choosing the appropriate basis to describe the different blocks turns out to be the choice that minimizes the information loss in the system as one increases the system size, and is precisely what is needed to produce an accurate and efficient algorithm. In a more modern language, we like to think of the density matrix renormalization group ansatz for the wavefunction as part of a general matrix product state ansatz\cite{oestlund}. Then, quantum information theory tells us that this approach is a form of a quantum compression algorithm\cite{vidal1,vidal2,verstraete_dmrg} and one can quantify the computational cost and the accuracy via studying the associated entanglement and entanglement entropy. Recent developments have also shown how the density-matrix renormalization group can be employed to benchmark density functional theory in one dimension\cite{burke}.

So why does the approach fail for higher dimensions? It turns out that the representation of the ground state in terms of the direct products with a truncated basis is an efficient way to expand wavefunctions that have limited entanglement.  As entanglement grows in the system, more complex ways to represent the wavefunction are needed, and generalizations from the matrix product state approach of the density matrix renormalization group to projected entangled pair states\cite{tensor}, and to multi-scale entanglement renormalization ansatzs\cite{mera}, show promise in being able to handle more entanglement in the wavefunction; they have already solved a number of difficult problems\cite{ipeps1,ipeps2,ipeps3}. In addition, progress has been made with solving long-range interactions, which can lead to efficient simulation of higher dimensional systems\cite{moore}.

There has been progress on solving time-dependent problems within the numerical
renormalization group framework\cite{anders}, but so far these problems have focused on quantum quenches, 
or steady-state properties of current flowing through a quantum dot attached to leads at two
different voltages. The basic idea to carry this out is that one can calculate the time-dependent expectation value of 
an operator from the time dependent density matrix.  Under the assumption that the system starts in a Hamiltonian $\mathcal{H}_0$ for early times, and is in equilibrium at some temperature, then the initial density matrix is just $\rho_0=\exp[{-\beta\mathcal{H}_0}]$. At time $t=0$ the system is
suddenly switched to a new Hamiltonian $\mathcal{H}_1$, and that Hamiltonian is responsible for all future time evolution of the system.  The time-dependent density matrix then satisfies
\begin{equation}
\rho(t)=e^{i\mathcal{H}_1t}\rho_0e^{-i\mathcal{H}_1t}.
\end{equation}
So the Wilson chain is constructed from the final Hamiltonian, and the evolution of the system is found in the long-time limit by averaging the time evolution, which corresponds to a dropping of the off-diagonal elements when expressed in an eigenbasis with respect to the final Hamiltonian. For current transport problems, it is important to formulate the system in terms of the scattering states of the noninteracting system, which is somewhat different from the conventional way of developing the numerical renormalization group because the impurity operators are part of the scattering states operators. Generalizations to include multiple quenches have also been made\cite{costi}. One of the issues with this approach is that the exponential decay of the hopping matrix elements projects the system to lower and lower energy shells of the final Hamiltonian, and it isn't clear that the nonequilibrium system is projected onto those low energy states. Hence, a new hybrid approach has been constructed where the numerical renormalization group approach is initially taken, and then one ends with a density matrix approach for the time-dependent problem\cite{nrg_dmrg}. This then allows for more accuracy at longer times when one does not continue to project onto lower energy shells.

The nonequilibrium dynamical mean-field theory problem requires a more complex approach to solving it because driving a system with a field, requires a more general nonequilibrium impurity solver than one that can handle quenches only. One has to first grapple with the so-called mapping problem, which involves determining the hybridization and the site-energy for each bath state that couples to the impurity\cite{eckstein_mapping}.  Unlike the construction for the real frequency case, where integration over the individual frequency intervals directly yielded the hybridization and bath energy, here, because of the more complex time dependence, one cannot easily determine the hybridization or bath energy. The problem has, however, been solved in principle, and it requires one to construct two sets of bath states, one called the $+$ bath and one the $-$ bath. The $+$ bath is constructed from the initial equilibrium mapping and the mixed real and imaginary time dynamical mean field. There is a unique and well defined strategy to determine these hybridization functions (which now depend on time).  Once they are found, they determine the hybridization function on the real axis.  But since this is most likely not the hybridization function of the nonequilibrium system, we need to subtract this result, and construct a second hybridization function and a second set of bath functions from the difference $\Lambda_-=\lambda-\Lambda_+$. Assuming this second dynamical mean field has a positive representation in terms of hybridization functions, one can then construct it.  One should imagine that the original imaginary axis dynamical mean field is given by the $\Lambda_+$ field, and that as time evolves, the initial correlations, given by the mixed real/imaginary dynamical mean field die off as the real time gets large, and the system evolves from the $\Lambda_+$ field to the $\Lambda_-$ field as time moves forward, as illustrated in Fig.~7. In any case, this approach has a constructive way to determine the effective hybridizations and bath energies, even though it is numerically challenging to solve these equations.

\begin{figure}[bt]
\centerline{\includegraphics[width=3.65in,clip]{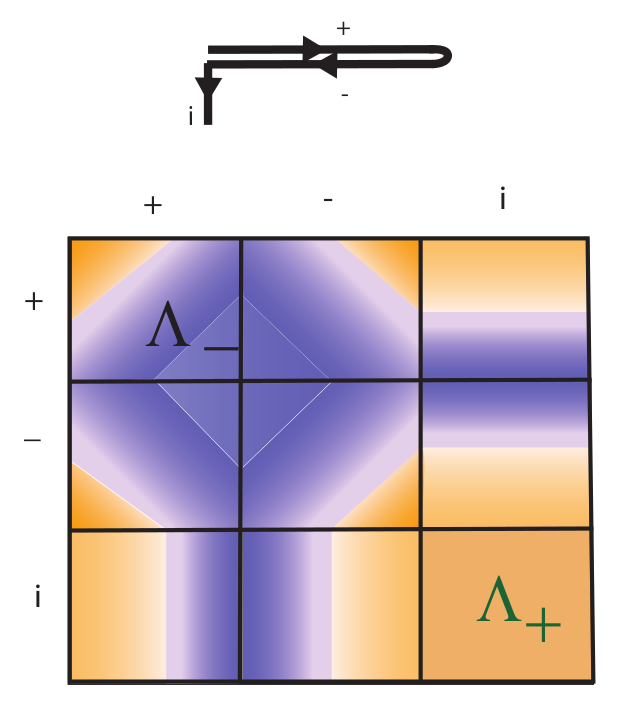}}
\vspace*{8pt}
\caption{(Color online) If we take the contour, shown at the top, and stretch it to a straight line, we can represent a matrix function that has each time lying on the contour on either the upper real branch (+), the lower real branch (-), or the imaginary branch ($i$). The $\Lambda_+$ field is largest in the equilibrium part of the matrix, closest to the imaginary spur, and fades away as one gets farther from that region (indicated by the orange color).  The other field $\Lambda_-$ grows the farther one gets from the spur  (indicated by the purple color). For large real times, the field is primarily $\Lambda_-$.}
\end{figure}

The problem for developing a complete nonequilibrium numerical renormalization group approach, if that even is possible, has not yet been completed. Since the driven system need not be restricted to a low energy subspace, nor do high energy degrees of freedom need to be frozen in at low temperature, the renormalization
aspect of the algorithm may not hold anymore. Numerical attempts have been made recently\cite{eckstein,schollwoch}, but more work needs to be done to make these approaches more efficient. These ideas provide an interesting starting point for algorithms that focus on the causal nature of the many-body problem and are described in more detail below.

We conclude this section be describing the time-dependent density matrix renormalization group approach\cite{marston,feigun}. Here, one wants to find the wavefunction as a function of time, and it will change in time, according to how the Hamiltonian evolves, by simply following the equation of motion. Since we consider a Hamiltonian that changes as a function of time, a simple Trotterization of the evolution operator is appropriate, and so the algorithm needs to be modified from calculating the ground state wavefunction to determining the action of $\exp[-i\mathcal{H}(t)\Delta t]$ on the wavefunction at time $t$. It turns out this can also be calculated using sparse matrix techniques based on either the Lanczos algorithm or the Crank-Nicolson algorithm. It is best to perform these calculations for the finite-system algorithm, because then one can zip to the left and then to the right for each time step and update the system accordingly.  Of course, the set of $m$ kept basis state functions will now change as a function of time, and they will continue to be determined by the reduced density matrix of the Hamiltonian at the current time instant. It is not obvious that this produces the optimal basis to work in, but it is how current calculations are performed. Once again, if the time evolution of the system creates significant entanglement growth, or if the disturbance in the system moves to the boundaries where it will reflect, one can no longer evolve the equations further in time to determine the properties in the long-time limit. In any case, this algorithm has proved to be one of the best approaches to the nonequilibrium problem in one dimensional systems, and it is likely that any improvements to the approach need to adopt a similar time evolution of the active subspace within which the current quantum state lies, as it evolves forwards in time.

There have been some interesting new developments that incorporate Chebyshev polynomials for expansions of the many-body density of states in equilibrium\cite{new_dmrg}.  The density of states is found by recursively determining moments of the Chebyshev expansion, which involves a matrix-vector multiplication to determine the recursion.  The full matrix is truncated by employing a matrix-product-state representation for the wavefunction used in the recursion relation.  It is possible ideas represented in this work will be useful also for nonequilibrium problems and for deciding on appropriate compression schemes for representing the data for the wavefunction.

\section{Algorithms based on the evaluating the evolution operator}

In our final technical section, we discuss the the so-called nonequilibrium Green's function approach, which works on effective quadratic Hamiltonians (which are either noninteracting, or describe mean-field couplings, including low-temperature superconductivity). Quadratic Hamiltonians have the benefit of being able to be diagonalized exactly, and hence, a calculation of the evolution operator can be performed exactly, and from the evolution operator, one can directly find the Green's function. (For example, the retarded Green's function is given by the evolution operator, while the lesser Green's function, which depends on the full history of the system, can be constructed from the evolution operator as well, but is a much more complicated object.) Hence, these approaches effectively are focused on determining the evolution operator directly.

The nonequilibrium Green's function approach\cite{meir,datta}, is used to solve for the nonequilibrium transport in mesoscopic systems that are connected to ideal (ballistic metal) leads.  The leads are maintained at fixed temperatures and voltages, and by projecting the Dyson equation onto the
active interacting part of the device, where the mesoscopic behavior occurs, one can describe the leads
via effective self-energy contributions determined by the surface Green's functions of the leads.  This then maps an infinite system onto an effective finite-sized system that can then be treated numerically. In the case where the many-body dynamics is restricted to a mean-field treatment, or is described by simple perturbation theory, one can immediately solve these problems, although the numerical effort can be huge, and one has to pay particular detail to conservation laws, gauge invariance, electronic charge reconstruction, and self-consistency in solving both the quantum mechanics and the electrodynamics of the driven system\cite{stefannuci,branislav_pt}.

Recently, there has been a breakthrough on the formal/numerical side by mapping the nonequilibrium Green's function approach onto a simple wavefunction based approach\cite{waital}.  If one decomposes the retarded Green's function into a sum over products of wavefunctions, and examines the equation of motion for the wavefunctions, one finds they satisfy the time-dependent Schr\"odinger equation, with the full time-dependent Hamiltonian, plus an additional source term that arises from the currents being driven by the leads. Since the Green's function is composed from the wavefunctions, essentially as the sum over outer products, one can reduce the complex matrix-based numerics for solving the Green's functions by vector-based numerics for solving the wavefunctions.  Here, one of the critical issues one needs to preserve in solving these problems is to maintain unitarity of the time evolution, which can rapidly be lost in conventional integrators of wavefunctions (or Green's functions), but there are algorithms to do this. The main challenge with this wave-function-based approach, is that it is not easily generalized to full many-body interacting problems, but it does hint at the fact that one can find much more efficient algorithms for solving these types of problems by determining the correct way to represent the Green's functions.  In particular, if it is possible to represent the full many-body problem in terms of a wavefunction-based representation with a simple Schr\"odinger-like equation with an additional source term, then one can extend these wavefunction-based methods much more broadly.

Motivated by the success of this approach, one can consider other methods which have not been exploited too much to try to tackle the problem of determining the evolution operator.  One method to consider is to factorize the evolution operator into a sequence of factors which might make it easier to evaluate the full operator numerically.  This approach is well known, as it forms the basis for the interaction-picture representation of the evolution operator, but often the factorization stops there.  In cases where the Hamiltonian does not commute with itself at different times, but its commutator does (i.e., $[\mathcal{H}(t),\mathcal{H}(t')]$ commutes with $\mathcal{H}(t'')$), one can find the exact evolution operator by breaking the Hamiltonian into a series of three factors, as is often used in solving for the driven harmonic oscillator in quantum mechanics textbooks\cite{landau,gottfried}.  By employing clever decompositions of the Hamiltonian, as is done with the density-matrix renormalization group method, one might find new ways to factorize the evolution operator that can make computation much simpler with it. We feel this has been an underexplored area for the nonequilibrium many-body problem.

\section{Computer Science ideas for extreme data science}

From a computer science perspective, one would like to determine an efficient way to encode 
the large complex data set, construct a small database of how that set evolves as a function of time, 
and use the current information to forecast future information. In principle, this should be possible, because the structure of the formula for the Green's function is a trace over initial states that have high thermodynamic weight.  At low temperatures, this is restricted to a small set of states, which becomes just the ground state at $T=0$. The evolution of this state through the Hilbert space traces out a one-dimensional path, and hence it always is a small subspace of the full space.  The problem arises when we try to represent this state in a particular basis. As the wavefunction becomes more entangled, or involves mixtures of more eigenstates, it can take a large number of states to represent it. One key element within this is to employ 
causality as the unifying principle for how to evolve the system forward in time. For the quantum many-body problem, 
this entails determining an equation of motion and solving it. One can do so by determining the evolution operator, or a
projection of the evolution operator onto the active subspace of the Hilbert space, or via a direct integration of the
wavefunction. To do either of these, requires an efficient way to encode the evolution operator or the wavefunction, for 
otherwise, the representation of these objects will grow too rapidly to be able represent it for any reasonable amount 
of time. Since the operation needed at any given time step is the action of the evolution operator on the particular wavefunction that one is following, it is likely that one can evolve the wavefunction efficiently.

The simplest method for encoding the basis vectors is to pick a single-particle basis and encode the state as a binary 
with a one indicating a particle in that state and a zero indicating no particle in that state. As the 
density-matrix renormalization group studies have shown, clever ways of encoding the wavefunction, via tensor products 
or other types of data structures, can prove to be much more efficient ways of encoding the wavefunction and allowing it 
to grow in complexity as a function of time. More research and more ideas of how to create, manipulate, and calculate 
with such data structures is imperative to make progress with finding efficient encoding schemes. It is also likely that one will need to change the bases and let them evolve as a function of time for more efficient implementations.  Clearly, computer 
science expertise in appropriate data structures will be helpful in making progress with this approach.

In addition to novel data structures, appropriate computing architectures and paradigms are needed to support extreme
data science computations. Dataflow and computing models that display extreme parallelism are among such possibilities to be considered\cite{data_flow}.  Dataflow computing arises from situations where
storing the entire data space is impossible (which encompasses extreme data science by definition), so the algorithms that direct computation are based on the 
availability and flow of the data and are able to employ parallelism at a high level. Developing 
this type of computation is the
 foundational approach to solving extreme data science problems.  

Furthermore, computing paradigms such as lazy computation\cite{lazy} and speculative computing\cite{speculative} must also be considered.  Lazy computation
delays the actual computation until absolutely necessary.  In our focus, such an approach supports the projection of
our global data space to only what is immediately needed and can be maintained (the time-evolving target data subspace).  Speculative computation supports simultaneous multiple path solution space
explorations providing the ability to sustain approximating boundary conditions as needed.

The most numerically challenging scheme will be to determine how to either solve for the evolution operator of the system, or how to integrate the wavefunction as a function of time. Ideas along these lines will be presented in the next section.

Forecasting falls into two realms.  One is extrapolating the average time to longer times and the other involves extrapolating the relative time to longer times.  It turns out that the latter is often much easier than the former to accomplish. For equilibrium problems, the time translation invariance of the dynamical mean field, often implies that the relative time dependence can be determined by the determinant of a large Toeplitz matrix\cite{freericks_toeplitz}. Using Szego's theorem and the elegant mathematics of the Wiener-Hopf approach\cite{wiener,krein,mccoy_wu}, one can construct an exact formula for the asymptotic exponential decay of the Green's function as a function of relative time. If the spectral function has power-law singularities, then the approach needs to be modified to the Fisher-Hartwig conjecture\cite{fisher}, which will determine the power-law decay of the Green's function in relative time. It is only in the unlikely case of a delta function in the spectral function, that one encounters a relative time dependence that does not decay. For the nonequilibrium problem, if sufficient data have been generated for long enough times, one can extract the appropriate decay forms for longer times, under the assumption that the decaying behavior as a function of relative time can be extrapolated from the short-time results. The expectation is that the Green's function always decays with a complex exponential behavior for large relative times.

Extrapolation as a function of average time requires some form of an ansatz for the behavior of the system at long times.  Since the retarded Green's function determines the quantum states of the system, it is expected to rapidly approach its long-time behavior, while the lesser Green's function, which determines how those states are filled, often takes much longer to reach its asymptotic limit.  But in cases where it is described by an effective fluctuation-dissipation theorem with a time-dependent temperature, the extrapolation can be carried out quite effectively\cite{fotso}. In other cases, if one knows the analytic behavior of the system, one can extrapolate summations of diagrams to capture the long-time limits as well\cite{nikolic}. Techniques along these lines have been used within numerical approaches for describing asymptotic expansions, but much more can be done by enforcing properties of analyticity, where they apply. They generically fall into the mathematical theory of resurgence\cite{ecalle}, which has been applied to quantum mechanics, quantum field theory and string theory\cite{pham,marino}, and has great potential to be applied to the nonequilibrium many-body problem.

It is also advantageous to use reduced dynamics techniques, which re-express long-time dynamics in terms of a short-time memory kernel, which generically decays much faster in time than the Green's function\cite{cohen1}. The short-term kernel can be determined by computational methods for nonequilibrium that are accurate for short times, and then can be input into the reduced dynamics formalism and evaluated in the long-time limit. This approach has been successfully applied to a number of different problems.  One example is the application to spin relaxation in the Kondo regime with the bold continuous time quantum Monte Carlo method\cite{cohen2}.

\section{Future thoughts and conclusions}

We examined the behavior of a number of different algorithms for calculating properties of many body systems in equilibrium and in nonequilibrium. Clearly, the nonequilibrium
problem is much more complicated than the equilibrium one, and it may require the development of new approaches to be able to reach long times with a general purpose algorithm. One result which remains clear, is that stochastic methods, which map the quantum problem onto an effective statistical problem, will suffer from the dynamical phase problems, which will greatly limit their ability to perform accurate calculations for long times.  This can only be fixed by finding alternative representations that can forestall the phase problem from occurring.  The bold method shows promise in this area, but still has much development to go before it will be useful for general problems. If we leave these stochastic methods off the table as we develop novel approaches that will likely work for nonequilibrium, we are left with focusing on either determining the evolution operator directly, as a function of time (and perhaps projected onto the active subspace), or evolving the wavefunction directly in time. These methods are closely related to exact diagonalization techniques and to density matrix renormalization group methods, which both use this approach. For dynamical mean-field theory applications, we always need to solve an effective single impurity Anderson model, but we believe that it must be solved in alternative ways than the conventional time-dependent numerical renormalization group, because one need not project onto an effective low energy space for a driven system. We call these time-driven methods, methods based on causality, since we will utilize the causal nature of the wavefunctions or Green's functions in solving the problem. Recent ideas along these lines can be found in an approach that focuses on the evolution of expectation values in the Heisenberg picture for the single impurity Anderson model\cite{cohen3}.

We want to discuss the application of the four computer science ideas within this many-body context.  First, we have the issue of how to encode the data. Here, the density matrix renormalization group is probably the most sophisticated approach, as it tries to create wavefunctions that can contain a certain amount of entanglement in their representation. Determining the most efficient way to represent the wavefunction is one of the unsolved goals, but there has been some new work along these lines in the context of the numerical renormalization group in equilibrium\cite{gunnarson} which could point in a new direction for how to formulate the solution of this problem.  One of the key elements is to focus on how to efficiently update the basis that is used from one time step to another, and new ideas within this context are certainly going to be needed.

The evolution of the system boils down to one of two techniques.  The first is to solve for the evolution operator (which can then subsequently be used to find the Green's functions). One approach that has not been used too much is to factorize the evolution operator.  While everyone learns the simplest factorization as seen with the interaction representation, it might be possible to factorize further via a Magnus-like expansion\cite{magnus}, where subsequent factorizations are possible if the potential in the interaction picture can be broken up into two pieces, whose commutator at two different times commutes with all operators. The second approach is to directly solve for the wavefunction as a function of time.  This approach, via a direct integration of the Schr\"odinger equation, is often the most efficient way to proceed, because it requires one to operate just the Hamiltonian on the wavefunction at each time instant, but many algorithms that do this suffer from loss of unitarity, and hence the algorithm needs to be designed with care to guarantee it will preserve the total probability.

None of the evolution methods in time will work without having an ability to truncate the results into an active subspace of the whole Hilbert space.  The density matrix renormalization group uses the reduced density matrix to determine this optimal basis, while the recent work on the numerical renormalization group by Gunnarsson and collaborators\cite{gunnarson} used a simple truncation scheme to trim the number of basis vectors used by their amplitude to appear in the wavefunction.  Both methods show great promise in finding ways to reduce the active subspace so that the problem does not grow to be too large as a function of time, and to adapt the target subspace as time moves forward. Efficient coding of the states will be needed to determine the full history of the evolution.

Finally, the ideas for forecasting in this problem rely strongly on the analytic properties of the functions being calculated, and by using as much of these analytic properties as possible, one can, perhaps, evolve these systems much further in time than one thought they could be evolved simply by extrapolating with the appropriate analytic functions for the long-time behavior. 

The basic ideas of extreme data science, the ability to work with data sets that are so large they cannot be constructed or stored on any computer, seems to be impossible to deal with at first glance.  By carefully analyzing these problems, especially for the class of problems where the active subspace remains small as the system evolves in time, it is clear that much progress can be made.  The nonequilibrium many-body problem can serve as a unique paradigm where a cross fertilization of ideas from physics, computer science and applied mathematics can help solve just such a problem and potentially lay the ground work for how one solves similar classes of problems in other fields. Doing so could revolutionize the field of big data science and also the scientific fields that can be successfully analyzed within these contexts.

\section*{Acknowledgments}

We acknowledge useful conversations with John Arrington, Guy Cohen, Peter Drummond, Adrian Feiguin, Ushma Kriplani, Hector Mera, Roman Orus,  Andreas Stathopoulos, and Xavier Waintal.
J. K. F. acknowledges the support of the Department of Energy Office of Basic Energy Science under grant number DE-FG02-08ER46542 and the McDevitt bequest at Georgetown University.
B. K. N. acknowledges support of the National Science Foundation under grant number
ECCS-1202069.
O. F. acknowledges the McDevitt bequest at Georgetown University.
J. K. F. also thanks the Aspen Center for Physics and the National Science Foundation under Grant Number 1066293 for hospitality during the initial stages of developing this review.

\end{document}